\documentclass[twocolumn,showpacs,aps,prd,amsmath,amssymb]{revtex4-1}

\usepackage[dvips]{graphicx}
\usepackage{subfigure}
\usepackage{color}
\usepackage{epsfig}
\usepackage{indentfirst}
\usepackage{multirow}
\usepackage{amsmath}
\usepackage{amssymb}
\usepackage{bm}
\usepackage{fancyhdr}
\usepackage{enumerate}
\usepackage[perpage,symbol]{footmisc}
\usepackage[bookmarksnumbered,pdfstartview=FitH,colorlinks=true,menucolor=black,linkcolor=blue]{hyperref}
\usepackage{xspace}
\begin{document}

\title{%
 \boldmath Measurement of the form factors in the decay $D^+ \to \omega e^+ \nu_{e}$
   and search for the decay $D^+ \to \phi e^+ \nu_{e}$}

\author{
  \begin{small}
    \begin{center}
      M.~Ablikim$^{1}$, M.~N.~Achasov$^{9,f}$, X.~C.~Ai$^{1}$,
      O.~Albayrak$^{5}$, M.~Albrecht$^{4}$, D.~J.~Ambrose$^{44}$,
      A.~Amoroso$^{49A,49C}$, F.~F.~An$^{1}$, Q.~An$^{46,a}$,
      J.~Z.~Bai$^{1}$
% ...
      \\
      \vspace{0.2cm}
      (BESIII Collaboration)\\
      \vspace{0.2cm} {\it
        $^{1}$ Institute of High Energy Physics, Beijing 100049, People's Republic of China\\
% ...
$^{4}$ Bochum Ruhr-University, D-44780 Bochum, Germany\\
        $^{5}$ Carnegie Mellon University, Pittsburgh, Pennsylvania 15213, USA\\
% ...
$^{9}$ G.I. Budker Institute of Nuclear Physics SB RAS (BINP), Novosibirsk 630090, Russia\\
% ...
$^{44}$ University of Rochester, Rochester, New York 14627, USA\\
% ...
$^{46}$ University of Science and Technology of China, Hefei 230026, People's Republic of China\\
% ...
$^{49}$ (A)University of Turin, I-10125, Turin, Italy; (B)University of Eastern Piedmont, I-15121, Alessandria, Italy; (C)INFN, I-10125, Turin, Italy\\
% ...
$^{a}$ Also at State Key Laboratory of Particle Detection and Electronics, Beijing 100049, Hefei 230026, People's Republic of China\\
% ...
$^{f}$ Also at the Novosibirsk State University, Novosibirsk, 630090, Russia\\
% ...
      }
    \end{center}
    \vspace{0.4cm}
  \end{small}
}
\affiliation{}

\date{\today}

\begin{abstract}

Using 2.92 fb$^{-1}$ of electron-positron annihilation data collected
at a center-of-mass energy of $\sqrt{s}= 3.773$ GeV
with the BESIII detector,  we present an improved measurement of the branching fraction
$\mathcal{B}(D^+ \to \omega e^+ \nu_{e}) = (1.63\pm0.11\pm0.08)\times 10^{-3}$.
The parameters defining the corresponding hadronic form factor ratios
at zero momentum transfer are determined for the first time; we
measure them to be $r_V = 1.24\pm0.09\pm0.06$
and $r_2 = 1.06\pm0.15 \pm 0.05$.
The first and second uncertainties are statistical and systematic, respectively.
We also search for the decay $D^+ \to \phi e^+ \nu_{e}$. An
improved upper limit $\mathcal{B}(D^+ \to \phi e^+ \nu_{e}) < 1.3 \times 10^{-5}$ is set at 90\% confidence level.
\end{abstract}


\maketitle
\pagenumbering{arabic}
\vspace{10mm}

Charm semileptonic decays have been studied
in detail because they provide essential inputs of the magnitudes of the
Cabibbo-Kobayashi-Maskawa (CKM) elements
$|V_{cd}|$ and $|V_{cs}|$~\cite{Cabibbo:1963yz,Kobayashi:1973fv},
and a stringent test of the strong interaction effects in the decay amplitude.
These effects of the strong interaction in
the hadronic current are parameterized by form factors that are calculable,
for example, by lattice QCD and QCD sum rules.
The couplings $|V_{cs}|$ and $|V_{cd}|$ are tightly
constrained by the unitarity of the CKM matrix.
Therefore, measurements of charm semileptonic decay rates and form
factors rigorously test theoretical predictions.
Both high-statistics and rare modes should be studied
for a comprehensive understanding of charm semileptonic decays.

\begin{figure}
\centering
  \includegraphics[width=0.35\textwidth]{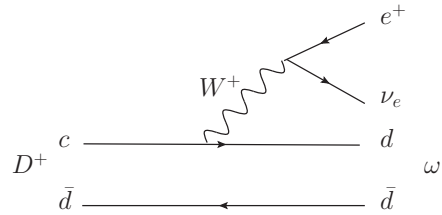}
  \caption{Feynman diagram representing the tree-level charged current
    process $D^+ \to \omega e^+ \nu_e$.}\label{fig:feynmanomega}
\end{figure}

For $D \to V \ell \nu$ transitions (where $V$ refers
to a vector meson),  the form factors have been studied
in the decays  $D^+ \to \overline{K}^{*0} e^+ \nu_e$~\cite{delAmoSanchez:2010fd}
and $D^+ \to \rho^0 e^+ \nu_e$~\cite{CLEO:2011ab}.  The decay
$D^+ \to \omega e^+ \nu_e$ was first observed by the CLEO-c
experiment,  while the corresponding form factors have not yet been
measured due to limited statistics~\cite{CLEO:2011ab}.
The decay $D^+ \to \omega e^+ \nu_e$ can proceed through the
tree-level diagram shown in
Fig.~\ref{fig:feynmanomega}. Its transition rate depends on
the charm-to-down-quark coupling $|V_{cd}|$, which is precisely known from
unitarity of the CKM matrix. Neglecting the lepton mass,
three dominant form factors contribute
to the decay rate: two axial ($A_1$,$A_2$) and one vector ($V$) form factor,
which are functions of the square of the invariant mass of the lepton-neutrino system $q^2$.

\begin{figure}
\centering
  \includegraphics[width=0.35\textwidth]{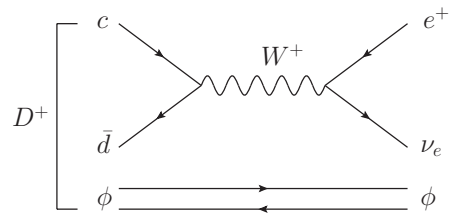}\\
  \caption{Feynman diagram representing the WA process $D^+ \to \phi e^+ \nu_e$.}\label{fig:feynmanphi}
\end{figure}

The decay $D^+ \to \phi e^+ \nu_e$ has not
yet been observed.  The most recent experimental search was performed by the
CLEO Collaboration in 2011 with a sample of an integrated luminosity
of 818 pb$^{-1}$ collected at the $\psi(3770)$ resonance.
The upper limit of the decay rate was set to be $9.0\times 10^{-5}$ at
the 90\% confidence level (C.L.)~\cite{Yelton:2010js}.
Since the valence quarks $s\bar{s}$ of the $\phi$ meson are distinct from those of
the $D$ meson ($c\bar{d}$), this process cannot occur in the absence of $\omega$-$\phi$
mixing or a non-perturbative ``weak annihilation'' (WA) contribution as shown
in Fig.~\ref{fig:feynmanphi}~\cite{Gronau:2009mp,Bigi:1980zpc}.
A measurement of the branching fraction can discriminate which process is dominant.
For example, a study of the ratio of $D^+_s \to \omega e^+ \nu_e$
and $D^+_s \to \phi e^+ \nu_e$~\cite{Gronau:2009mp}
concludes that any value of $\mathcal{B}(D^+_s \to \omega e^+ \nu_e)$ exceeding
$2\times 10^{-4}$ is unlikely to be attributed to $\omega$-$\phi$ mixing, and would
provide evidence for non-perturbative WA effects~\cite{Bigi:1980zpc}.
A search for the decay $D^+ \to \phi e^+ \nu_e$ is helpful, since its
dynamics is similar to that of the decay $D^+_s \to \omega e^+ \nu_e$.

In this paper, we report an improved measurement of $\mathcal{B}(D^+ \to \omega e^+ \nu_e)$ and
the first form factor measurement in this decay. Furthermore, an improved upper limit
for $\mathcal{B}(D^+ \to \phi e^+ \nu_e)$ is determined. Charge conjugate states are implied throughout
this paper. Those decays are studied using a data sample collected with the BESIII detector
which corresponds to an integrated luminosity of 2.92 fb$^{-1}$ at the $\psi(3770)$ resonance~\cite{BESIII:2013iaa}.

The BESIII detector is a spectrometer located at BEPCII, which is a double-ring
$e^+ e^-$ collider working at the center-of-mass energy range from 2 to 4.6 GeV.
The cylindrical core of the BESIII detector consists of a
helium-based multi-layer drift chamber (MDC), a plastic scintillator
time-of-flight system (TOF), and a CsI (Tl) electromagnetic
calorimeter (EMC), which are all enclosed in a superconducting
solenoid magnet providing a 1.0 T magnetic field.  The solenoid is
supported by an octagonal flux-return yoke with modules of resistive
plate muon counters interleaved with steel. The
momentum resolution for charged particles at 1 GeV/$c$ is 0.5\%, and
the resolution of the ionization energy loss per unit path-length ($dE/dx$)
is 6\%. The EMC measures photon energies with
a resolution of 2.5\% (5\%) at 1 GeV in the barrel (end cap). The
time resolution of the TOF is 80 ps in the barrel and 110 ps in the
end cap. A detailed description of the BESIII detector
is provided in Ref.~\cite{Ablikim:2009aa}.

The tagging technique for the branching fraction measurements
of semileptonic decays was first employed by the
Mark-III collaboration~\cite{Adler:1989rw} and later applied in
the studies by CLEO-c~\cite{CLEO:2011ab,Huang:2005iv}.
The presence of a $D^+ D^-$ pair in an event
allows a \emph{tag sample} to be defined in which a $D^-$ is
reconstructed in one of the
following six hadronic decay modes: $K^+ \pi^- \pi^-$,
$K^+ \pi^- \pi^- \pi^0$, $K^0_S \pi^-$, $K^0_S \pi^- \pi^0$,
$K^0_S \pi^+ \pi^- \pi^-$, and $K^+ K^- \pi^-$.
A sub-sample is then defined in which a positron and a set of hadrons
are required recoiling against the tag $D$ meson, as a signature of a semileptonic decay.
The yields of tag and signal are expressed as
$N_{\mathrm{tag}}^i = 2 N_{D\bar{D}}\mathcal{B}_{\mathrm{tag}}^i\epsilon_{\mathrm{tag}}^i$ and
$N_{\mathrm{sig}}^i = 2 N_{D\bar{D}}\mathcal{B}_{\mathrm{tag}}^i \mathcal{B}_{\mathrm{sl}} \epsilon_{\mathrm{tag,sl}}^i$,
respectively, where $N_{D\bar{D}}$ is the total number of $D\bar{D}$ pairs produced, $i$
indicates a tag mode, $N_{\mathrm{tag}}^i$ is the number of observed
tag events in mode $i$,
$N_{\mathrm{sig}}^i$ is the number of semileptonic candidates,
$\mathcal{B}_{\mathrm{tag}}^i$ is the branching fraction of tag mode $i$,
$\mathcal{B}_{\mathrm{sl}}$ is the branching fraction of the semileptonic decay,
$\epsilon_{\mathrm{tag}}^i$ is the reconstruction efficiency of a tag
mode, and $\epsilon_{\mathrm{tag,sl}}^i$ is the reconstruction efficiency
of the semileptonic decay with a tag mode. Thus, $\mathcal{B}_{\mathrm{sl}}$ can be expressed as
\begin{equation}\label{equ:bsig}
\centering
\mathcal{B}_{\mathrm{sl}} = \frac{N_{\mathrm{sig}}}{\displaystyle{\sum_i} N_{\mathrm{tag}}^i \epsilon_{\mathrm{tag,sl}}^i /\epsilon_{\mathrm{tag}}^i} \ ,
\end{equation}
where $N_{\mathrm{sig}}$ is the total signal yield in all six tag modes.

Charged tracks are reconstructed using hit information from the MDC.
The tracks are required to satisfy $|\cos\theta| < 0.93$, where $\theta$
is the polar angle with respect to the beam axis.
Tracks (except for $K^0_S$ daughters) are required to
originate from the interaction point (IP), \emph{i.e.} their point of
closest approach to the interaction point is required to be $\pm 10$ cm along the beam direction and
$1$ cm transverse to the beam direction.
Charged particle identification (PID) is accomplished
by combining the $dE/dx$ and TOF information to form a likelihood $\mathcal{L}_i$
($i=e/\pi/K$) for each particle hypothesis. A $K^{\pm}$ ($\pi^{\pm}$)
candidate is required to satisfy $\mathcal{L}_K > \mathcal{L}_{\pi}$
($\mathcal{L}_{\pi} > \mathcal{L}_{K}$). For electrons,
we require the track candidate to satisfy
$\frac{\mathcal{L}_{e}}{\mathcal{L}_{e}+\mathcal{L}_{\pi}+\mathcal{L}_{K}} > 0.8$
as well as $E/p \in [0.8,1.2]$, where $E/p$ is the ratio
of the energy deposited in the EMC to the momentum of the track
measured in the MDC. In order to take into account the effect of
final state radiation and bremsstrahlung, the energy of neutral clusters
within $5^\circ$ of the initial electron direction is assigned
to the electron track.
The $K^0_S$ candidates are reconstructed from pairs of oppositely charged
tracks, which are assumed to be pions and required to have
an invariant mass in the range $m_{\pi^+\pi^-} \in [0.487,0.511] \ \mathrm{GeV}/c^{2}$.
For each pair of tracks, a vertex-constrained fit is performed to ensure
that they come from a common vertex.

To identify photon candidates, showers must have minimum energies of 25 MeV
in the barrel region ($|\cos\theta|<0.80$) or 50 MeV in the end cap region
($0.86<|\cos\theta|<0.92$). To exclude showers from
charged particles, a photon candidate must be separated by at least
20$^\circ$ from any charged track with respect to the IP.
A requirement on the EMC timing
suppresses electronic noise and energy deposits unrelated to the event.
The $\pi^0$ candidates are reconstructed from pairs of photon
candidates by requiring the invariant
di-photon mass to fulfill $m_{\gamma \gamma}\in [0.115, 0.150] \ \mathrm{GeV}/c^{2}$.
Candidates with both photons coming from the end cap region are rejected due to poor resolution.

The $D^-$ tag candidates  are selected based on two variables: $\Delta E \equiv E_D - E_{\mathrm{beam}}$,
the difference between the energy of the $D^-$ tag candidate ($E_D$) and
the beam energy ($E_{\mathrm{beam}}$), and the beam-constrained
mass $M_{\mathrm{bc}} \equiv \sqrt{E^2_{\mathrm{beam}}/c^4 - |\vec{p}_D|^2/c^2}$, where
$\vec{p}_D$ is the measured momentum of the $D^-$ candidate.
In each event, we accept at most one candidate per tag mode per charge,
and the candidate with the smallest $|\Delta E|$ is chosen.
The yield of each tag mode is obtained from fits to the $M_{\mathrm{bc}}$ distributions~\cite{Ablikim:2013uvu}.
The data sample comprises about $1.6\times 10^6$ reconstructed charged tag
candidates (Table~\ref{tab:dtagdata}).

\begin{table}
\centering
  \caption{Tag yields of the $D^-$ six hadronic modes and their statistical uncertainties.} \label{tab:dtagdata}
\begin{tabular}{lc}
\hline \hline
Tag mode                &   $N_{\mathrm{tag}}^i$ \\
\hline
    $D^- \to K^+ \pi^- \pi^-$ & $809425\pm906$ \\
    $D^- \to K^+ \pi^- \pi^- \pi^0$ & $242406\pm599$ \\
    $D^- \to K^0_S \pi^-$ & $100149\pm321$ \\
    $D^- \to K^0_S \pi^- \pi^0$ & $226734\pm575$ \\
    $D^- \to K^0_S \pi^+ \pi^- \pi^-$ & $132683\pm489$ \\
    $D^- \to K^+ K^- \pi^-$ &     $70530\pm325$ \\
    \hline
    Total           &  \quad  \quad $1581927\pm1399$ \quad  \quad \\
    \hline \hline
  \end{tabular}
\end{table}

Once a $D^-$ tag candidate is identified, we search for an $e^+$ candidate and an $\omega \to \pi^+ \pi^- \pi^0$
candidate or a $\phi \to K^+ K^-$ candidate recoiling against the tag.
If there are multiple $\omega$ candidates in an event, only
one combination is chosen based on the proximity of the $\pi^+ \pi^- \pi^0$  invariant
mass to the nominal $\omega$ mass~\cite{Beringer:1900zz}.
The invariant mass $m_{\pi^+ \pi^- \pi^0}
\in [0.700,0.840]\ \mathrm{GeV}/c^{2}$ and $m_{K^+ K^-} \in [1.005,1.040]\
\mathrm{GeV}/c^{2}$ are required for $\omega$ and $\phi$ candidates,
which correspond to three times of the $\omega$ ($\phi$)
mass resolution ($\pm 3\sigma$), respectively.
For the decay $D^+ \to \omega e^+ \nu_e$, backgrounds arise mostly from $D^+ \to \overline{K}^{*0} e^+ \nu_e$,
$\overline{K}^{*0} \to K_S^0 \pi^0$, $K_S^0 \to \pi^+ \pi^-$ process, and
the invariant mass of the charged pions is required to be outside
the aforementioned $K_S^0$ mass region.
This requirement rejects about $70\%$ of the $D^+ \to
\overline{K}^{*0} e^+ \nu_e$ background events.

After tag and semileptonic candidates have been combined, all charged tracks
in an event must be accounted for.
The total energy of additional photon candidates, besides those used in the tag and semileptonic
candidates, is required to be less than $0.250\ \mathrm{GeV}$.
Semileptonic decays are identified using the variable $U \equiv E_{\mathrm{miss}} - c|\vec{p}_{\mathrm{miss}}|$,
where $E_{\mathrm{miss}}$ and $\vec{p}_{\mathrm{miss}}$ are the missing energy and
momentum corresponding to the undetected neutrino from the $D^+$ meson semileptonic decay, which are calculated by
$E_{\mathrm{miss}} \equiv E_{\mathrm{beam}} - E_{\omega(\phi)} - E_{e}$,
$\vec{p}_{\mathrm{miss}} \equiv - (\vec{p}_{\mathrm{tag}} + \vec{p}_{\omega(\phi)} + \vec{p}_{e})$
in the center-of-mass frame, where $E_{\omega(\phi)}$ ($E_{e}$) and
$\vec{p}_{\omega(\phi)}$ ($\vec{p}_{e}$) are the energy and momentum
of the hadron (electron) candidate. To obtain a better $U$ resolution,
the momentum of the tag $D^-$ candidate $\vec{p}_{\mathrm{tag}}$ is
calculated by $\vec{p}_{\mathrm{tag}} = \widehat{p}_{\mathrm{tag}}[(E_{\mathrm{beam}}/c)^2-M_D^2c^2]^{1/2}$~\cite{Coan:2005iu},
where $\widehat{p}_{\mathrm{tag}}$ is the unit vector in the direction of the tag $D^-$
momentum, and $M_D$ is the world average value of $D$ meson mass~\cite{Beringer:1900zz}.
The correctly reconstructed semileptonic candidates are expected to
peak around zero in the $U$ distribution.
A {\sc geant4}-based~\cite{Agostinelli:2002hh} Monte Carlo (MC) simulation
is employed, and events are generated with {\sc kkmc}+{\sc evtgen}~\cite{kkmc,evtgen}
to determine the efficiencies in Eq.~(\ref{equ:bsig}), as
shown in Table~\ref{tab:eff}.
All selection criteria and signal regions are defined using simulated
events only.

\begin{table}
\centering
\caption{Tag efficiencies ($\epsilon_{\mathrm{tag}}$) and signal efficiencies
including a tag ($\epsilon_{\mathrm{tag,sl}}$) in percent, as
determined by the MC simulation, and their statistical uncertainties.}
\label{tab:eff}
\begin{tabular}{lccc}
\hline \hline
Tag mode                         & $\epsilon_{\mathrm{tag}}$  & $\epsilon_{\mathrm{tag,sl}}$ ($\omega$) & $\epsilon_{\mathrm{tag,sl}}$ ($\phi$) \\
\hline
$D^- \to K^+ \pi^- \pi^-$        & $51.07\pm0.02$    &  $11.22\pm0.10$                &   $9.04\pm0.09$ \\
$D^- \to K^+ \pi^- \pi^- \pi^0$  & $25.13\pm0.02$    &  $5.15\pm0.09$                 &   $4.38\pm0.08$ \\
$D^- \to K^0_S \pi^-$            & $54.40\pm0.05$    &  $11.70\pm0.32$                &   $9.69\pm0.29$ \\
$D^- \to K^0_S \pi^- \pi^0$      & $29.24\pm0.02$    &  $6.13\pm 0.11$                &   $5.34\pm0.10$ \\
$D^- \to K^0_S \pi^+ \pi^- \pi^-$& $37.61\pm0.04$    &  $7.28\pm0.18$                 &   $5.96\pm0.16$ \\
$D^- \to K^+ K^- \pi^-$          & $41.12\pm0.06$    &  $8.97\pm0.29$                 &   $7.63\pm0.27$ \\
\hline \hline
\end{tabular}
\end{table}

\begin{figure}
\centering
  \includegraphics[width=0.4\textwidth]{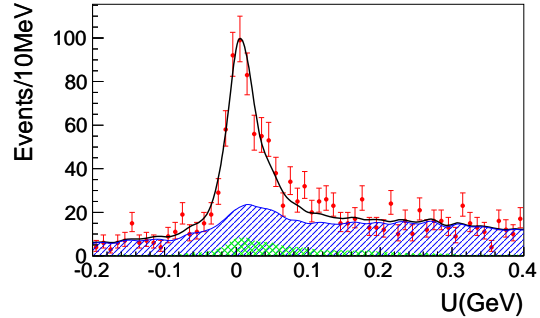}\\
  \caption{Fit (solid line) to the $U$ distribution in data (points with error bars)
  for the semileptonic decay $D^+ \to \omega e^+ \nu_e$. The total background contribution
  is shown by the filled curve, while the peaking component is shown
  by the cross-hatched curve.}\label{fig:omegadatau}
\end{figure}

The yield of the decay $D^+ \to \omega e^+ \nu_e$ is obtained from a fit to
the $U$ distribution combining all tag modes, as shown in Fig.~\ref{fig:omegadatau}.
The signal shape is described by the shape from the signal MC simulation convoluted
with a Gaussian function whose width is left free in the fit
to describe the resolution difference between MC and data.
The background model consists of two components: peaking and non-peaking backgrounds.
Peaking background arises mostly from the decay $D^+ \to \overline{K}^{*0} e^+ \nu_e$,
$\overline{K}^{*0} \to K_S^0 \pi^0$, $K_S^0 \to \pi^+ \pi^-$; its $U$ distribution
is modeled with MC simulation.
The largest contribution to the non-peaking backgrounds is from
the $D\bar{D}$ process, while the remaining background events are from the non-$D\bar{D}$,
$q\bar{q}$, $\tau^+\tau^-$, initial state radiation $\gamma J/\psi$ and
$\gamma \psi(2S)$ processes. The non-peaking component is modeled with
a smooth shape obtained from MC simulations.
In the fit to data, the yield of
the peaking background is fixed to the MC expectation, while that of the non-peaking
background is left free in the fit. The signal yield is determined by
the fit to be $N_{\mathrm{sig}} =
491\pm32$. The absolute branching fraction of the decay $D^+ \to \omega e^+ \nu_e$ as listed in
Table~\ref{tab:bf} is obtained using Eq.~(\ref{equ:bsig}).

\begin{figure}
\centering
  \includegraphics[width=0.4\textwidth]{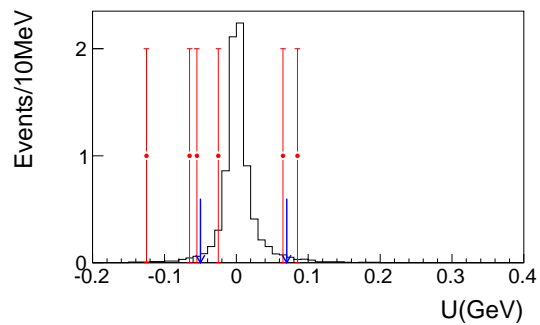}\\
  \caption{The $U$ distribution for the semileptonic decay $D^+ \to \phi e^+ \nu_e$ in
    data (points with error bars) and signal MC simulation with
    arbitrary normalization (solid histograms). The arrows show
    the signal region.}\label{fig:phidatau}
\end{figure}

The $U$ distribution for the decay $D^+ \to \phi e^+ \nu_e$ with all tag
modes combined is shown in Fig.~\ref{fig:phidatau}. The signal region is defined as
$[-0.05,0.07]\ \mathrm{GeV}$, which covers more than $97\%$ of all signal events according
to MC simulations. No significant excess of signal events is observed, and there are only 2
events in the signal region. A simulation study indicates that the backgrounds arise mostly
from $D^+ \to \phi \pi^+ \pi^0$ and $D^+ \to \phi \pi^+$ processes. The number of
background events is estimated to be $4.2 \pm 1.5$ via large
statistics MC samples.
The upper limit is calculated by using a frequentist method with
unbounded profile likelihood treatment of systematic uncertainties,
which is implemented by a C++ class {\sc TROLKE} in the ROOT
framework\cite{Rolke:2004mj}. The number of the observed events
is assumed to follow a Poisson distribution, and the number of
background events and the efficiency are assumed to follow Gaussian
distributions.
The resulting upper limit on $\mathcal{B}(D^+ \to \phi e^+ \nu_e)$ at $90\%$ C.L. is
determined to be $1.3 \times 10^{-5}$, as shown in
Table~\ref{tab:bf}.

With the double tag technique, the branching fraction measurements are insensitive to
systematics from the tag side since these are mostly cancelled.
For the signal side, the following sources of systematic uncertainty are taken into account, as
summarized in Table~\ref{tab:susum}. The uncertainties of tracking
and $K^{\pm}/\pi^{\pm}$ PID efficiencies are well studied by double
tagging $D\bar{D}$ hadronic decay events. The uncertainties in $e^{\pm}$
tracking and PID efficiency are estimated with radiative Bhabha events.
The uncertainty due to the $\pi^0$ reconstruction
efficiency is estimated with a control sample $D^0 \to K^- \pi^+ \pi^0$
by the missing mass technique. The uncertainty
due to imperfect knowledge of the semileptonic
form factors is estimated by varying the form factors in the MC
simulation according to the uncertainties on the measured
form factor ratios in the decay $D^+ \to \omega e^+ \nu_e$
as discussed below.  For the decay $D^+ \to \phi e^+ \nu_e$,
the signal MC produces phase-space distributed events, and therefore uses a constant form
factor. To evaluate the corresponding systematics, the form factor is varied by a
reweighting technique~\cite{Martin:2011rd}. The world average values of
$\mathcal{B}(\omega \to \pi^+ \pi^- \pi^0)$ and $\mathcal{B}(\phi \to K^+ K^-)$
are $(89.2\pm0.7)\%$ and $(48.9\pm0.5)\%$, respectively, and their uncertainties are
assigned as systematic uncertainties due to the input branching fractions in the MC simulation.
The limited MC statistics also leads to a systematic uncertainty.
The uncertainties associated with the $\omega$ and $\phi$ mass requirements are estimated using
the control samples $D^0 \to \omega K^- \pi^+$ and $D^+ \to \phi \pi^+$, respectively.
The $K^0_S$ rejection leads to an uncertainty on the signal efficiency of the decay $D^+ \to \omega e^+ \nu_e$,
which is studied by the control sample $D^0 \to \omega K^- \pi^+$.
The uncertainty due to the extra shower veto is studied with double hadronic tags.
For the decay $D^+ \to \phi e^+ \nu_e$, the uncertainty due to the
signal window requirement as shown in Fig.~\ref{fig:phidatau} is
estimated by the control sample
$D^+ \to \overline{K}^{*0} e^+ \nu_e$, $\overline{K}^{*0} \to K^- \pi^+$.
In the fit to the $U$ distribution in the $D^+ \to \omega e^+ \nu_e$ decay,
the uncertainty due to the parametrisation of the signal shape is estimated by varying the
signal shape to a Crystal Ball function~\cite{cbf}.
The uncertainty due to the fit range is estimated by varying
the fit range.
The uncertainty due to the non-peaking background is estimated
by modeling this component with a third-order Chebychev function,
and the uncertainty associated with the fixed peaking
background normalization is estimated by varying it within its expected uncertainty.
All of those estimates are added in quadrature to obtain the total systematic
uncertainties on the branching fractions.


\begin{table}
\begin{footnotesize}
\caption{Measured branching fractions in this paper and a comparison
to the previous measurements~\cite{CLEO:2011ab,Yelton:2010js}. For the
$D^+ \to \omega e^+ \nu_e$ decay,
the first uncertainty is statistical and the second systematic.}\label{tab:bf}
\centering
\begin{tabular}{lcc}
\hline \hline
Mode                &     This work                   &     Previous      \\
\hline \hline
$\omega e^+ \nu_e$    &     $(1.63\pm0.11\pm0.08)\times 10^{-3}$           &     $(1.82\pm0.18\pm0.07) \times 10^{-3}$ \\
\hline
$ \phi e^+ \nu_e$   &  $< 1.3 \times 10^{-5}$ ($90\%$C.L.)    &      $< 9.0 \times 10^{-5}$ ($90\%$C.L.)   \\
\hline \hline
\end{tabular}
\end{footnotesize}
\end{table}

\begin{table}
  \caption{Summary of systematic uncertainties on the branching fraction
  measurements.}\label{tab:susum}
  \begin{tabular}{lcc}
    \hline \hline
    Source                 &    $\mathcal{B}(D^+ \to \omega e^+ \nu_e)$   &   $\mathcal{B}(D^+ \to \phi e^+ \nu_e)$  \\
    \hline
    Tracking           &        $3.0\%$                             &       $3.0\%$                           \\
    $K$/$\pi$ PID      &        $1.0\%$                             &       $1.0\%$                           \\
    $e$ PID           &         $3.2\%$                            &        $3.4\%$                          \\
    $\pi^0$ reconstruction        &         $1.0\%$                &           -                             \\
    Model of form factor &      $1.0\%$                            &        $1.2\%$                          \\
    $\omega$($\phi$) decay rate  &         $0.8\%$                            &        $1.0\%$                          \\
    MC statistics          &         $0.7\%$                            &        $0.9\%$                          \\
    $\omega$($\phi$) mass window  &         $0.9\%$                            &         $0.4\%$                         \\
    $K^0_S$ veto           &         $0.2\%$                            &           -                             \\
    Extra shower veto      &         $0.1\%$                            &          $0.1\%$                         \\
    Signal window          &            -                               &          $0.4\%$                         \\
    Fit range          &         $0.4\%$                            &            -                             \\
    Signal shape        &         $0.6\%$                            &            -                             \\
    Peaking background     &         $0.8\%$                            &             -                            \\
    Non-peaking background &          $0.4\%$                           &             -                            \\
    \hline
    Total                  &             $5.1\%$                         &          $5.0\%$                         \\
    \hline \hline
  \end{tabular}
\end{table}

\begin{figure}
\centering
\includegraphics[width=0.4\textwidth]{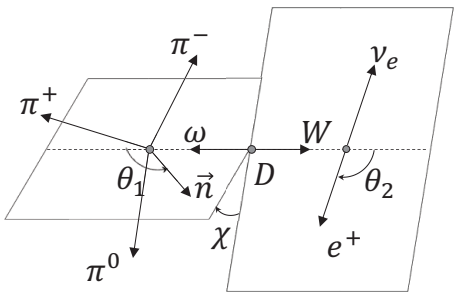}\\
\caption{Definitions of the helicity angles in the decay $D^+ \to \omega W^+$, $\omega \to \pi^+ \pi^- \pi^0$, $W^+ \to e^+ \nu_e$ for
  the three-body ($\theta_1$) and two-body ($\theta_2$) $D^+$-daughter decays, where
  both angles are defined in the rest-frame of the decaying meson.}
\label{fig:defination}
\end{figure}

The differential decay rate of $D^+ \to \omega e^+ \nu_e$ can be expressed in
the following variables as illustrated in Fig.~\ref{fig:defination}:
$m^2$, the mass square of the $\pi\pi\pi$ system; $q^2$, the mass
square of the $e\nu_e$ system;
$\theta_1$, the $\omega$ helicity angle~\cite{Berman:1965pr}, which
is the angle between the $\omega$ decay plane normal ($\widehat{n}$)
in the $\pi\pi\pi$ rest frame and the direction of flight of the $\omega$ in the
$D$ rest frame;
$\theta_2$, the helicity angle of $e$, which is the angle
between the charged lepton three-momentum in the $e\nu_e$
rest frame and the direction of flight of the $e\nu_e$ system in the $D$ rest frame;
 $\chi$, the angle between the decay planes of those two systems.

For the differential partial decay width, only the $P$-wave component is taken into consideration
and the formalism expressed in terms of three helicity amplitudes
$H_+(q^2)$, $H_-(q^2)$, and $H_0(q^2)$ is~\cite{Korner:1989qb,Gilman:1989uy,CLEO:2011ab}:
\begin{widetext}
\begin{equation}
    \begin{split}
    \frac{d\Gamma}{dq^2 d\cos\theta_1 d\cos\theta_2 d\chi dm_{\pi\pi\pi}}  =  & \frac{3}{8(4\pi)^4} G^2_F |V_{cd}|^2 \frac{p_\omega q^2}{M_{D}^2} \mathcal{B}(\omega\to\pi\pi\pi)|\mathcal{BW}(m_{\pi\pi\pi})|^2 \\
    & [(1+\cos\theta_2)^2 \sin^2\theta_1 |H_+(q^2,m_{\pi\pi\pi})|^2 \\
    & + (1-\cos\theta_2)^2 \sin^2\theta_1 |H_-(q^2,m_{\pi\pi\pi})|^2 + 4 \sin^2 \theta_2 \cos^2 \theta_1 |H_0(q^2,m_{\pi\pi\pi})|^2 \\
    & + 4 \sin \theta_2 (1+\cos\theta_2)\sin \theta_1 \cos \theta_1 \cos \chi H_+(q^2,m_{\pi\pi\pi}) H_0(q^2,m_{\pi\pi\pi}) \\
    & - 4 \sin \theta_2 (1-\cos\theta_2)\sin \theta_1 \cos \theta_1 \cos \chi H_-(q^2,m_{\pi\pi\pi}) H_0(q^2,m_{\pi\pi\pi}) \\
    & - 2 \sin^2 \theta_2 \sin^2 \theta_1 \cos 2\chi H_+(q^2,m_{\pi\pi\pi}) H_-(q^2,m_{\pi\pi\pi})].
    \end{split}
    \label{eq:amplitude}
\end{equation}
\end{widetext}
where $G_F$ is the Fermi constant, $p_\omega$ is the momentum of the $\omega$
in the $D$ rest frame, $\mathcal{B}(\omega\to\pi\pi\pi)$ is the branching fraction of $\omega \to \pi\pi\pi$,
$m_{\pi\pi\pi}$ is the invariant mass of the three pions, and $\mathcal{BW}(m_{\pi\pi\pi})$
is the Breit-Wigner function that describes the $\omega$ line shape.
The helicity amplitudes can in turn be related to the two axial-vector form
factors $A_{1,2}(q^2)$ and the vector form factor $V(q^2)$:
\begin{equation}
\begin{split}
H_{\pm}(q^2) =  & M A_1(q^2) \mp 2 \frac{M_D p_{\omega}}{M}V(q^2) \\
H_0(q^2)     =  & \frac{1}{2m_{\pi\pi\pi}\sqrt{q^2}}[(M_D^2-m_{\pi\pi\pi}^2-q^2)M A_1(q^2)\\
             &  -4\frac{M_D^2 p_{\omega}^2}{M} A_2(q^2)]
\end{split}
\label{eq:hpm0def}
\end{equation}
where $M=M_D + m_{\pi\pi\pi}$.
For the $q^2$ dependence, a single pole parameterization~\cite{Richman:1995wm}
is applied:
\begin{equation}
V(q^2) =  \frac{V(0)}{1-q^2/m_V^2} \ ,\
A_{1,2}(q^2) =  \frac{A_{1,2}(0)}{1-q^2/m_A^2} \ ,
\label{eq:vadef}
\end{equation}
where the pole masses $m_V$ and $m_A$ are expected to be close to
$M_{D^*(1^-)}=2.01~\mathrm{GeV}/c^2$ and
$M_{D^*(1^+)}=2.42~\mathrm{GeV}/c^2$~\cite{Beringer:1900zz} for the vector and axial form
factors, respectively. The ratios of these form factors, evaluated at
$q^2 = 0$, $r_V = \frac{V(0)}{A_1(0)}$ and $r_2 = \frac{A_2(0)}{A_1(0)}$,
are measured in this paper.

\begin{figure}
\centering
\includegraphics[width=0.95\textwidth]{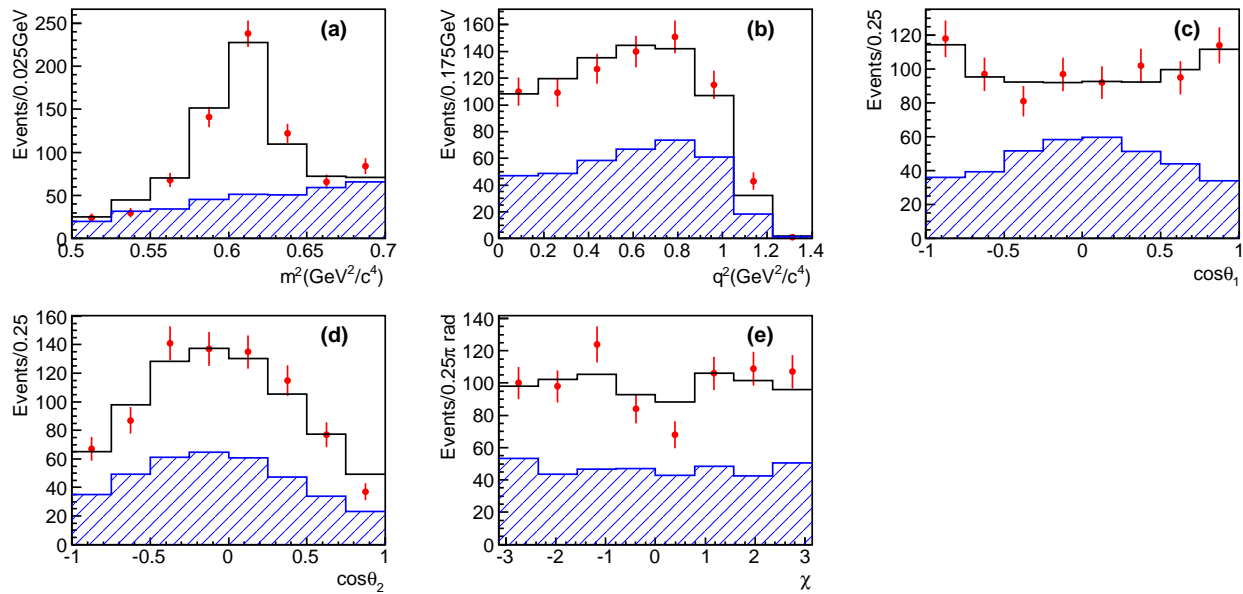}\\
\caption{Projections of the data set (points with error bars),
the fit results (solid histograms) and the sum of the background distributions
(filled histogram curves) onto (a) $m^2$, (b) $q^2$, (c) $\cos\theta_1$,
(d) $\cos\theta_2$ and (e) $\chi$.}
\label{fig:ffprojection}
\end{figure}

According to the fit procedure introduced in Ref.~\cite{delAmoSanchez:2010fd},
a five-dimensional maximum likelihood fit is performed in the space of $m^2$,
$q^2$, $\cos\theta_1$, $\cos\theta_2$ and $\chi$. The signal probability
density function is modeled with the phase-space signal MC events reweighted with
the decay rate (Eq.~\ref{eq:amplitude}) in an
iterative procedure.
The experimental acceptance is taken in consideration using this technique.
Large signal MC samples
are generated to reduce the systematic uncertainty associated with the
MC statistics. The background is modeled
with the MC simulation and its normalization
is fixed to the expectation.
Using simulated events with known $r_V$ and $r_A$, we verify that
this procedure can reliably determine the form factor ratios.
Figure ~\ref{fig:ffprojection} shows the $m^2$, $q^2$, $\cos\theta_1$, $\cos\theta_2$
and $\chi$ projections from the final fit to data. The fit determines
the form factor ratios to be $r_V = 1.24\pm0.09$ and $r_2 = 1.06\pm0.15$.

\begin{table}
\caption{Summary of the absolute systematic uncertainties in the form factor measurement
of the decay $D^+ \to \omega e^+ \nu_e$.}\label{tab:ffsusum}
\centering
\begin{tabular}{lcc}
\hline \hline
Sources                      &    \quad   $r_V$  \quad    & \quad    $r_2$  \quad    \\
\hline
$q^2$ dependence             &   \quad    0.05  \quad     &  \quad    0.03  \quad    \\
Background model          &  \quad     0.02 \quad      &  \quad    0.02  \quad    \\
Pole mass assumption         &   \quad    0.01  \quad     &    negligible     \\
Fitting shift                &    \quad   0.02  \quad     &  \quad    0.02  \quad    \\
\hline
Total                        &    \quad   0.06 \quad      &  \quad    0.05   \quad   \\
\hline \hline
\end{tabular}
\end{table}

For the form factor measurement in the decay $D^+ \to \omega e^+ \nu_e$,
the following sources of systematic uncertainties are taken into account, as
summarized in Table~\ref{tab:ffsusum}:
The uncertainty associated with the unknown $q^2$ dependence of the form factors is
estimated by introducing a double pole parameterization~\cite{Fajfer:2005ug}.
The uncertainty due to the background model is estimated by varying the
background normalization with its statistical uncertainty.
No events from the non-resonant decay $D^+ \to \pi^+ \pi^- \pi^0 e^+
\nu_e$ are observed, the influence of this decay on the form factor therefore can be neglected.
To estimate the uncertainty associated with
the pole mass assumption, we vary the pole mass $m_V$ by $\pm100\ \mathrm{MeV}/c^{2}$.
A small shift is observed with the presence of
background, and this is treated as possible bias in the form factor fitting
procedure. Adding all systematic uncertainties in quadrature,
the form factor ratios are determined to be $r_V = 1.24\pm 0.09 \pm 0.06$
and $r_2 = 1.06 \pm 0.15 \pm 0.05$, respectively.

In summary, using 2.92 fb$^{-1}$ of $e^+e^-$ annihilation data collected
at $\sqrt{s}=3.773$ GeV with the BESIII detector, we have measured
the form factor ratios in the decay $D^+ \to \omega e^+ \nu_e$
at $q^2=0$ for the first time: $r_V =  \frac{V(0)}{A_1(0)} = 1.24\pm0.09\pm0.06$,
$r_2 = \frac{A_2(0)}{A_1(0)} = 1.06\pm0.15 \pm 0.05$, and determined the branching
fraction to be $\mathcal{B}(D^+ \to \omega e^+ \nu_{e}) = (1.63\pm0.11\pm0.08) \times 10^{-3}$,
where the first and the second uncertainties are statistical and systematic, respectively.
This is the most precise measurement to date.
We have also searched for the rare decay $D^+ \to \phi e^+ \nu_{e}$ and
observe no significant signal.  We set an upper limit of
$\mathcal{B}(D^+ \to \phi e^+ \nu_{e}) < 1.3 \times 10^{-5}$ at the $90\%$ C.L.,
which improves the upper limit previously obtained by the CLEO
Collaboration~\cite{Yelton:2010js} by a factor of about 7.

The BESIII collaboration thanks the staff of BEPCII and the IHEP
computing center for their strong support. This work is supported in
part by National Key Basic Research Program of China under Contract
Nos. 2009CB825204, 2015CB856700; National Natural Science Foundation of China (NSFC)
under Contracts Nos.~10935007, 11125525, 11235011, 11322544, 11335008, 11425524;
the Chinese Academy of Sciences (CAS) Large-Scale Scientific Facility
Program; the CAS Center for Excellence in Particle Physics (CCEPP);
the Collaborative Innovation Center for Particles and Interactions
(CICPI); Joint Large-Scale Scientific Facility Funds of the NSFC and
CAS under Contracts Nos. 11179007, 11179014, U1232201, U1332201; CAS under
Contracts Nos. KJCX2-YW-N29, KJCX2-YW-N45; 100 Talents Program of CAS;
National 1000 Talents Program of China; INPAC and Shanghai Key
Laboratory for Particle Physics and Cosmology; German Research
Foundation DFG under Contract No. Collaborative Research Center
CRC-1044; Istituto Nazionale di Fisica Nucleare, Italy; Ministry of
Development of Turkey under Contract No. DPT2006K-120470; Russian
Foundation for Basic Research under Contract No. 14-07-91152; The
Swedish Resarch Council; U.S.~Department of Energy under Contracts
Nos. DE-FG02-04ER41291, DE-FG02-05ER41374, DE-FG02-94ER40823,
DESC0010118; U.S.~National Science Foundation; University of Groningen
(RuG) and the Helmholtzzentrum fuer Schwerionenforschung GmbH (GSI),
Darmstadt; WCU Program of National Research Foundation of Korea under
Contract No. R32-2008-000-10155-0.

\onecolumngrid

\twocolumngrid


\begin{thebibliography}{99}

\bibitem{Cabibbo:1963yz}
  N.~Cabibbo,
  Phys.\ Rev.\ Lett.\  {\bf 10}, 531 (1963).

\bibitem{Kobayashi:1973fv}
  M.~Kobayashi and T.~Maskawa,
  Prog.\ Theor.\ Phys.\  {\bf 49}, 652 (1973).

\bibitem{delAmoSanchez:2010fd}
  P.~del Amo Sanchez {\it et al.}  [BaBar Collaboration],
  Phys.\ Rev.\ D {\bf 83}, 072001 (2011).

\bibitem{CLEO:2011ab}
  S.~Dobbs {\it et al.}  [CLEO Collaboration],
  Phys.\ Rev.\ Lett.\  {\bf 110}, 131802 (2013).

\bibitem{Yelton:2010js}
  J.~Yelton {\it et al.}  [CLEO Collaboration],
  Phys.\ Rev.\ D {\bf 84}, 032001 (2011).

\bibitem{Gronau:2009mp}
  M.~Gronau and J.~L.~Rosner,
  Phys.\ Rev.\ D {\bf 79}, 074006 (2009).

\bibitem{Bigi:1980zpc}
   I.~I.~Bigi and N.~G.~Uraltsev, Nucl.\ Phys.\ B {\bf 423}, 33 (1994);
   H.~Y.~Cheng, Eur.\ Phys.\ J.\ C {\bf 26}, 551 (2003).

\bibitem{Martin:2011rd}
  L.~Martin {\it et al.}  [CLEO Collaboration],
  Phys.\ Rev.\ D {\bf 84}, 012005 (2011).

\bibitem{BESIII:2013iaa}
  M.~Ablikim {\it et al.} [BESIII Collaboration],
  Chin.\ Phys.\ C {\bf 37}, 123001 (2013).

\bibitem{Ablikim:2009aa}
  M.~Ablikim {\it et al.}  [BESIII Collaboration],
  Nucl.\ Instrum.\ Meth.\ A {\bf 614}, 345 (2010).

\bibitem{Adler:1989rw}
  J.~Adler {\it et al.}  [MARK-III Collaboration],
  Phys.\ Rev.\ Lett.\  {\bf 62}, 1821 (1989).

\bibitem{Huang:2005iv}
  G.~S.~Huang {\it et al.}  [CLEO Collaboration],
  Phys.\ Rev.\ Lett.\  {\bf 95}, 181801 (2005).

\bibitem{Ablikim:2013uvu}
  M.~Ablikim {\it et al.}  [BESIII Collaboration],
  Phys.\ Rev.\ D {\bf 89}, 051104 (2014).




\bibitem{Beringer:1900zz}
  K.A. Olive {\it et al.} [Particle Data Group], Chin. Phys. C, {\bf 38}, 090001 (2014).

\bibitem{Coan:2005iu}
  T.~E.~Coan {\it et al.}  [CLEO Collaboration],
  Phys.\ Rev.\ Lett.\  {\bf 95}, 181802 (2005).

\bibitem{Agostinelli:2002hh}
  S.~Agostinelli {\it et al.}  [GEANT4 Collaboration],
  Nucl.\ Instrum.\ Meth.\ A {\bf 506}, 250 (2003).

\bibitem{kkmc}
  S.~Jadach, B.~F.~L.~Ward and Z.~Was, Comp. Phys. Commu. {\bf 130}, 260 (2000);
  S.~Jadach, B.~F.~L.~Ward and Z.~Was, Phys. Rev. D {\bf 63}, 113009 (2001).

\bibitem{evtgen}
  D.~J.~Lange, Nucl. Instrum. Meth. A {\bf 462},152 (2001).


\bibitem{Rolke:2004mj}
  W.~A.~Rolke, A.~M.~Lopez and J.~Conrad,
  Nucl.\ Instrum.\ Meth.\ A {\bf 551}, 493 (2005).

\bibitem{cbf}
 J.~Gaiser. Ph.D. thesis, Stanford University Report No.~SLAC-255, 1982;
 T.~Skwarnicki, Ph.D. thesis, Jagiellonian University in Krakow DESY Report No.~F31-86-02, 1986.

\bibitem{Berman:1965pr}
  S.~M.~Berman and M.~Jacob, Phys.\ Rev.\ {\bf 139}, B1023 (1965).

\bibitem{Korner:1989qb}
  J.~G.~Korner and G.~A.~Schuler,
  Z.\ Phys.\ C {\bf 46}, 93 (1990).

\bibitem{Gilman:1989uy}
  F.~J.~Gilman and R.~L.~Singleton,
  Phys.\ Rev.\ D {\bf 41}, 142 (1990).

\bibitem{Richman:1995wm}
  J.~D.~Richman and P.~R.~Burchat,
  Rev.\ Mod.\ Phys.\  {\bf 67}, 893 (1995).

\bibitem{Fajfer:2005ug}
  S.~Fajfer and J.~F.~Kamenik,
  Phys.\ Rev.\ D {\bf 72}, 034029 (2005).

\end{thebibliography}
\end{document}